# Quantum secured gigabit optical access networks

Bernd Fröhlich[1,*], James F Dynes[1], Marco Lucamarini[1], Andrew W Sharpe[1], Simon W-B Tam[1], Zhiliang Yuan[1] & Andrew J Shields[1]

[1]Toshiba Research Europe Ltd, 208 Cambridge Science Park, Cambridge CB4 0GZ, UK
* bernd.frohlich@crl.toshiba.co.uk

**Optical access networks connect multiple endpoints to a common network node via shared fibre infrastructure. They will play a vital role to scale up the number of users in quantum key distribution (QKD) networks. However, the presence of power splitters in the commonly used passive network architecture makes successful transmission of weak quantum signals challenging. This is especially true if QKD and data signals are multiplexed in the passive network. The splitter introduces an imbalance between quantum signal and Raman noise, which can prevent the recovery of the quantum signal completely. Here we introduce a method to overcome this limitation and demonstrate coexistence of multi-user QKD and full power data traffic from a gigabit passive optical network (GPON). The dual feeder implementation is compatible with standard GPON architectures and can support up to 128 users, highlighting that quantum protected GPON networks could be commonplace in the future.**

# Introduction

Multiplexing of QKD[1,2] and strong classical data signals is an essential requirement for a seamless integration of QKD into existing telecommunication infrastructure. Whereas dedicated dark fibre links permit transmission of quantum states over long distances[3,4] and in harsh field environments[5-8], the presence of classical data signals in a live fibre makes the retrieval of quantum information more difficult due to excess noise generated by inelastic Raman scattering[9,10]. Progress has been made on operating QKD over point-to-point links carrying high-speed data signals[11,12]. However, realising multiplexed point-to-multipoint links critically depends on new approaches to overcome the heightened influence of Raman noise in passive networks. In these networks power splitters are employed to address each user. In a 128-user network the splitter adds at least 21 dB of extra optical loss to the quantum channel, while leaving the noise floor due to Raman scattering unattenuated. Although existing mitigation techniques like time and wavelength filtering or power control of the classical signals[13] strongly reduce the scattering noise, additional methods must be developed to integrate QKD in large scale networks.

Integration of QKD in GPON networks[14,15] is an appealing idea. In GPON networks eavesdropping is always possible in downstream direction, as the transmitted data is broadcast to all users. Every user can, in principle, intercept all downstream traffic[16]. This realistic threat has been acknowledged in the International Telecommunication Union standard developed for GPON, which supports encryption of the downstream broadcast[17].

The keys used for encryption, however, are not secure even against simple attacks[14]. QKD permits to close this security loophole, providing the encryption keys with information-theoretic security.

In this article we demonstrate a method to integrate multi-user QKD[18-21] in a GPON network which supports large scale networks. We first show that Raman noise normally restricts the achievable network capacity strongly. Then this limitation is overcome with the dual feeder architecture. In a dual feeder network, secure keys can be transmitted alongside full power GPON data signals without the need for post-processing[22] or time-alignment[23], making it compatible with this widespread optical access network technology. The method permits operating QKD with up to 128 users in realistic network layouts while keeping the advantage of having single fibre links in the main, multi-user part of the network.

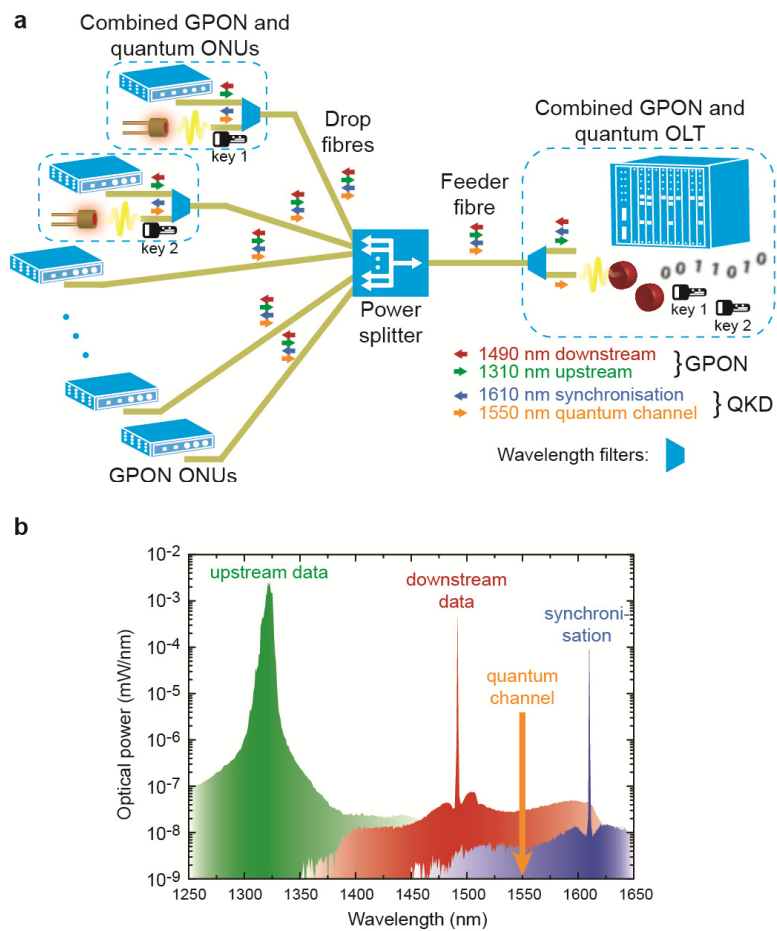

**Figure 1 Quantum secured optical access network. a**, In a passive optical network multiple users (ONU: optical network unit) are connected via drop fibres, an optical power splitter, and a feeder fibre to a network node (OLT: optical line terminal). We integrate QKD into the network using wavelength filters (trapezoid symbols). Quantum transmitters are installed in the ONUs and a shared quantum receiver is installed in the OLT. Each user exchanges individual encryption keys with the network node. **b**, Spectrum measured in upstream direction by inserting a 50:50 beam splitter in front of the OLT in an 8-user network. The spectrum shows peaks at 1310 nm and 1490 nm from data signals and a peak at 1610 nm from the synchronisation signal. The quantum signal at 1550 nm is completely obscured by the broad Raman scattering background.

# Results

## *Single feeder networks*

Figure 1a illustrates the combined quantum and classical network layout with a single feeder fibre between optical line terminal (OLT) and power splitter. For the quantum communication we employ an upstream quantum access network structure[19], where quantum transmitters are placed at the network endpoints (ONU: optical network unit) and the single-photon detector is situated at the network node. In addition to offering the advantage of sharing the complex single photon detector[24,25] between multiple users, it also permits an adjustable key transmission bandwidth per user. Each quantum transmitter sends individual keys encoded onto the phase of weak laser pulses transmitted at a rate of 1 GHz / $N$, where $N$ is the maximum number of users the network supports given by the splitting ratio of the passive splitter. Pulses from different transmitters are interleaved such that only photons from one transmitter arrive at the quantum receiver at any given time. For synchronisation the quantum receiver broadcasts a 250 MHz optical clock signal downstream to all users. The clock signal at 1610 nm and the quantum channel at 1550 nm are multiplexed with the GPON signals at 1310 nm (upstream) and 1490 nm (downstream). We employ standard telecom wavelength filters for this (see methods).

The classical signals are visible in the spectrum shown in Fig. 1b, which was measured in upstream direction by inserting a 50:50 beam splitter in front of the OLT in an 8-user network as shown in Fig. 1a. We use a single 15.5 km feeder fibre and two drop fibres of 4.4 km length for this measurement. Two combined quantum and GPON ONUs are connected to these drop fibres and 6 further GPON ONUs directly to the remaining ports of the power splitter. The downstream GPON signal is broadcast with full launch power of 4 dBm and a constant file transfer between all 8 GPON users guarantees a sustained total transmission power of the time-division multiplexed upstream signals of approximately 1 dBm. The upstream data is transmitted with inexpensive Fabry-Perot laser diodes generating the broad peak at 1310 nm. The spectrum also shows peaks from Rayleigh back-scattering at 1490 nm for the downstream data signal and at 1610 nm for the synchronisation signal. The broad Raman scattering background completely obscuring the quantum signal at 1550 nm shows that the majority of light scattered into the quantum wavelength band comes from the downstream GPON signal. It is transmitted with full power in the feeder fibre and its back-scattered photons are not attenuated by the splitter unlike the quantum signal.

To investigate the limitation imposed by back-scattering in the feeder fibre in more detail, we measure the secure key rate per quantum transmitter in this configuration for various feeder fibre lengths and launch powers. We keep the total fibre distance between quantum ONUs and OLT at 20 km and only change the ratio of feeder fibre $F$ to drop fibre $D$. The two quantum transmitters operate at 125 MHz. For more experimental details see the methods section. Figure 2a shows experimental and simulated data for the secure key rate per transmitter as a function of the single feeder fibre length $F$. The data illustrates that in networks with low capacity, for example for 8 users, successful key exchange over all distances is possible if the launch power is reduced strongly. A launch power of −11 dBm corresponds to a 3 dB margin to the point where the GPON network stops working (see methods). For higher power the secure key rate drops and eventually no key distillation is possible for all considered feeder fibre lengths. The problem, however, is much more severe if the splitting ratio is increased. Only very short feeder fibre lengths are supported in a 32-user network even if the power is reduced close to the lower limit.

The fundamental nature of this limitation is further illustrated in Fig. 2b. It shows a contour plot of QBER as a function of splitting ratio and feeder fibre length. The simulation takes only Raman noise in the feeder fibre into account as a source of error, neglecting, for example, detector and modulation imperfections. It therefore represents ideal conditions and normally the QBER will be higher than in the simulation. We assume an average photon flux of 0.5 photons per pulse, a total fibre distance of $F + D = 20$ km between OLT and ONUs, and a Fourier limited time-bandwidth product[11] of $\Delta v \cdot \Delta t = 0.44$ for wavelength filter width $\Delta v$ and detection gate width $\Delta t$. We assume an ideal splitter with insertion loss of $1 / N$ and the Raman scattering coefficient is set to $\beta_{DS} = 7.1 \times 10^{-9}$ nm$^{-1}$. The downstream launch power is adjusted to maintain a received power of –30 dBm which is 5dB below the receiver sensitivity specified for GPON class A optics[17] (see also methods). Even under these strongly idealised conditions QKD cannot be performed over the entire simulated range as the QBER crosses the threshold of 11% for the 4-state BB84 protocol[26].

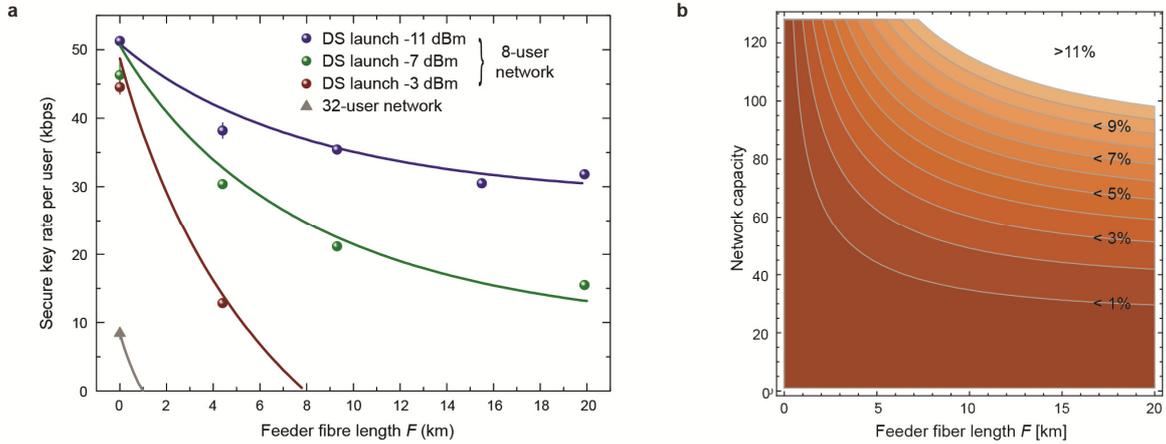

**Figure 2 Single feeder fibre network. a**, Secure key rate for the first quantum transmitter as a function of feeder fibre length $F$ in an 8-user single feeder network for different downstream (DS) data signal launch power. The total distance $F + D$ is kept equal to 20 km. Key transmission over the full length of 20 km is possible only for strongly attenuated downstream power. Shown in grey is the dependence for a 32-user network. Error bars correspond to 1 standard deviation of 3 consecutive measurements. The solid line is calculated using the numerical simulation described in the methods section. **b**, Simulation of QBER as a function of the feeder fibre length $F$ and splitting ratio in an idealised multiplexed quantum access network. The simulation takes only Raman photons in the feeder fibre into account as a source of errors.

## *Dual feeder networks*

We overcome this limitation by installing a $2 \times N$ passive optical splitter and a second feeder fibre as illustrated in the inset of Fig. 3a. In this configuration we re-measure the dependence of secure key rate on the length of the feeder fibre. All GPON signals are transmitted with full power in this experiment (see methods). Figure 3a summarises results for an 8-user network with a 2 × 8 passive splitter. It shows that the secure key rate is lowest if no feeder fibre is installed and the full distance is covered with the drop fibres. This result is intuitive as almost all Raman noise originates from the drop fibres in a dual feeder network. Because we keep $F + D$ constant, reducing $F$ corresponds to increasing $D$ and therefore the Raman noise, which increases with distance for short fibre lengths[13]. The decrease of the secure key rate of approximately 40% without feeder fibre compared to the data shown in Fig. 2 is due the higher launch power of the downstream GPON signal.

Having demonstrated that the situation with $F = 0$ always provides the lowest key rate, we set up the experiment in this worst-case configuration and increase the splitting ratio of the passive splitter. This conservatively bounds the maximum number of users addressable in a quantum secured GPON with two feeder fibres. Note that although we limit the number of GPON ONUs to 8, this is not restricting the amount of Raman noise generated by the upstream signal as the total upstream launch power does not increase with the number of GPON users due to time-division multiplexing. The effect of more quantum users and more fibre on the drop side of the network will be discussed below. Figure 3b displays how the key rate per transmitter depends on the network capacity. The secure key rate remains positive up to a capacity of 128. Therefore, we show compatibility with an even greater network size than demonstrated previously without multiplexing of GPON signals[19] which is mainly due to improvements to the single photon detector[27]. The total loss of fibre, splitter and filtering in this configuration is close to 30 dB which is equivalent to 150 km of standard single mode fibre with an attenuation of 0.2 dB/km. The average secure key rate per user is 0.5 kbps.

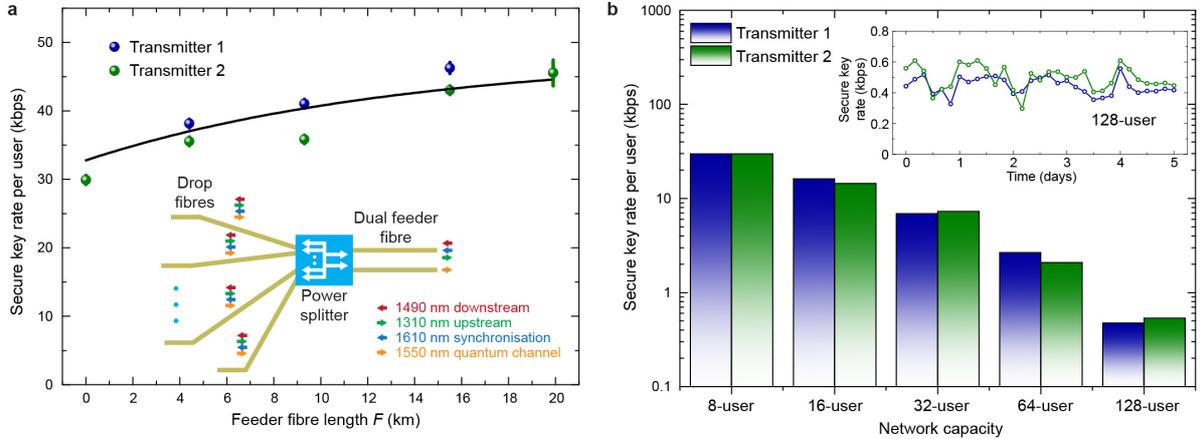

**Figure 3 Dual feeder fibre access network. a**, Secure key rate per quantum transmitter as a function of feeder fibre distance $F$ in a dual feeder network. The total distance $F + D$ is kept equal to 20 km. Error bars correspond to 1 standard deviation of 3 consecutive measurements. The solid line is calculated using the numerical simulation described in the methods section. Inset: Schematic of the dual feeder network. The power splitter is replaced with a 2 × $N$ splitter connected to two separate feeder fibres. The downstream GPON and synchronisation signal are launched into one feeder fibre, whereas the quantum signals is extracted from the second feeder fibre. **b**, Secure key rate per transmitter for varying network capacity with two feeder fibres. Secure transmission is demonstrated up to a splitting ratio of 2 × 128. Inset: Secure key rate over several days in a 128-user network.

## *Simulation*

The numerical simulation shown in Fig. 2 and 3 is based on a noise estimation using experimentally measured parameters (see methods). The noise estimation is calibrated through comparison with experimental data which permits us to investigate the effect of more quantum or classical users in the network on the key rate. Figure 4 shows simulation data of how the secure key rate per user changes when more users are added to the quantum access network. The network configuration corresponds to the experimental setup used to measure the data in Fig. 3b, i.e., no feeder fibre and 20 km drop fibre. We plot the secure key rate for four different network capacities of 16, 32, 64, and 128 users with each quantum transmitter operating at a speed of 1 GHz / N. The key rates therefore are lower by at least a factor of $N / 8$ compared to the data presented in Fig. 3b due to the lower operational speed. We keep the key session times to the values reported in the methods section, which enhances the influence of the finite sample size accordingly. Additionally, more users lead to more afterpulsing noise as the total detection count rate increases[19]. Nevertheless, the secure key rate stays positive for all network capacities as the data shows.

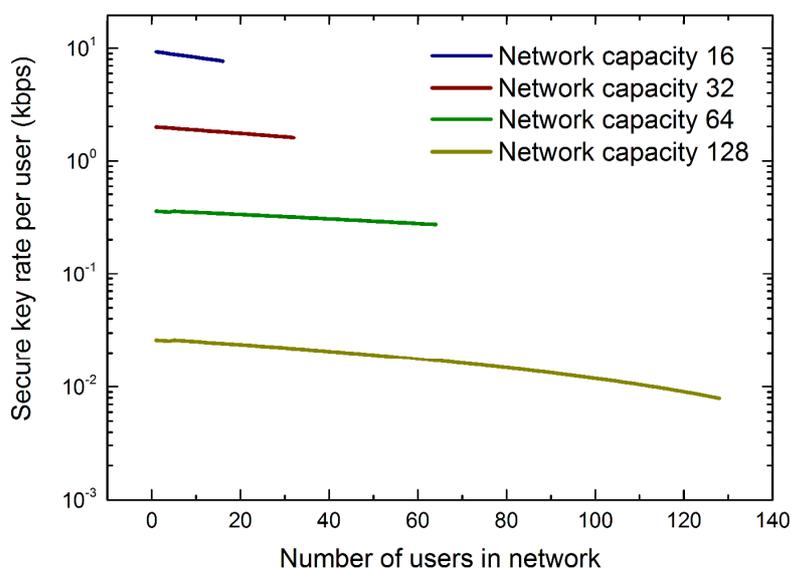

**Figure 4 Secure key rate in a quantum network with more users.** The secure key rate decreases when more users are added to the network but stays positive for all network capacities considered. The reduction stems from higher afterpulsing noise in a network where more of the detection bandwidth is used[19]. The key rates are overall reduced compared to the data shown Fig. 3 due to the lower operational speed of the transmitters (see text).

In a dual feeder network almost all Raman noise originates from the drop fibres. The QBER therefore increases if more drop fibre is added between ONUs and the passive splitter. Figure 5 shows how the secure key rate changes as a function of total drop fibre length in the network. For the simulation we set the feeder fibre length to zero. We assume a full quantum network with each transmitter operating at 1 GHz / $N$ and a full GPON network with $N$ GPON ONUs. Here, we increase the key session time to compensate for the reduced key size in a quantum network with more users. Note that methods such as trusting the detector[3] or adding more decoy states[28] could be used instead to increase the resilience against statistical fluctuations. For all considered network capacities the secure key rate stays positive for drop fibre lengths of well above $N \times 10$ km, therefore showing that quantum secured GPONs are viable in a large range of different network architectures. Note that typically drop fibres are shorter than 10 km and instead more distance is covered by the feeder fibre, which will result in higher secure key rates.

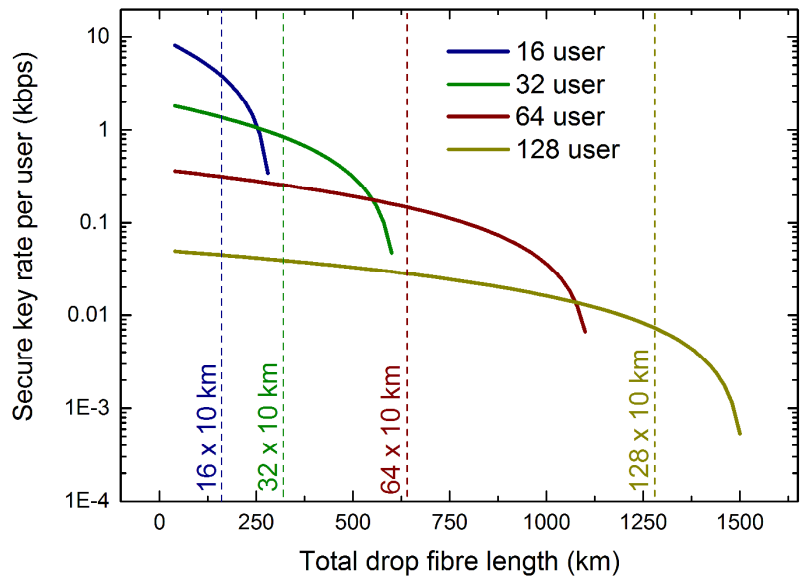

**Figure 5 Secure key rate as a function of total drop fibre.** Simulated data for a full quantum and full GPON network with varying capacity. Due to increasing Raman noise the secure key rate decreases with the amount of drop fibre in the network. It stays positive, however, up to total drop fibre lengths of more than $N \times 10$ km for all capacities indicated by the dashed lines. More drop fibre can be tolerated in larger capacity networks because the optical power transmitted per ONU decreases with the number of ONUs due to time-division multiplexing.

## Discussion

We have demonstrated that QKD can be implemented in typical GPON networks employing two feeder fibres between the network node and passive splitter. The GPON data signals are transmitted at full power permitting a smooth upgrade to a quantum secured network. Note, that a similar method would not be viable in a downstream quantum access network configuration[18] where dual drop fibres would be required. Two separate feeder fibres are used in conventional GPON networks for PON protection[17]. If one link fails, the network switches to the second link and therefore reduces the downtime of the system. This protection scheme is more likely to be integrated in critical network infrastructure, where QKD could be of particular benefit[29]. PON protection is also included in the next generation 10 gigabit GPON standard[30], making the method compatible with the next generation of optical access networks.

# Methods

## Quantum access network

The QKD network consists of two quantum transmitters exchanging secure keys simultaneously with the shared quantum receiver using time-division multiplexing. Each transmitter implements a phase encoding efficient BB84 protocol with decoy states and biased basis choice, guaranteeing composable security in the finite size scenario[31]. Attenuated laser pulses with 30ps full width at half maximum (FWHM) are modulated to signal, decoy, and vacuum state levels chosen with probabilities of 85.7%, 9.5%, and 4.8%, respectively. The respective intensities are 0.49, 0.03, and 0.0005 photons/pulse. We narrow the spectrum of the pulses with a 25 GHz dense wavelength division multiplexing (DWDM) filter based on a fibre Bragg grating to a FWHM of approximately 18 GHz to match a second filter in the receiver. The pulse width is broadened to 60 ps in this step. A phase modulator inside an asymmetric Mach-Zehnder interferometer encodes bit and basis. We select the minority basis with a probability of 1/16, our protocol is therefore strongly biased. The quantum receiver consists of a matching interferometer and two self-differencing InGaAs/InP single photon detectors[27] operated at 1 GHz and –30° C with efficiencies of approximately 26%, dark count probability of $2 \times 10^{-6}$, and afterpulse probability of 2% for a dead time of 300 ns. The transmitters are synchronised with the receiver with a 250 MHz optical clock signal. The power of this signal is 75 µW in the 8-user network and it is adjusted to compensate for the increasing splitter loss in networks with larger capacity. The transmitters operate at 125 MHz and are time aligned such that photon pulses arrive at the receiver separated by 4 detection gates. Polarisation and phase control is implemented in each transmitter using count rate and QBER as feedback signal.

## GPON and Raman scattering

The conventional access network consists of 8 ONUs and the OLT. For ease of use we implemented Ethernet passive optical network (EPON) which in its physical implementation is similar to the ITU standard for gigabit-capable passive optical networks[17]. The network operates with a speed of 1.25 Gbps in both upstream and downstream direction. The maximum launch power of the 1490 nm downstream data signal is 4 dBm. The power of the upstream signal depends on how much of the bandwidth of the network is used. We measure an average transmission power per ONU of about –8 dBm at 1310 nm while sustaining full network usage by running a constant file transfer between all ONUs. When attenuating the downstream signal the network stops working for an ONU receive power of approximately –30 dBm. This is 5 dB below the receiver sensitivity specified for class A optics in the GPON standard[17] and permits to reduce the launch power below the level required for a standard receiver. Forward Raman scattering of the upstream signal is given by[13]

$$P_f = P_{US} \beta_{US} \Delta\lambda \Delta t \frac{\left(e^{-\alpha_q L} - e^{-\alpha_{US} L}\right)}{\alpha_{US} - \alpha_q} ,$$

where $P_{US}$ is the launch power, $\alpha_q$ and $\alpha_{US}$ are the fibre attenuation coefficients for 1550 nm and 1310 nm, respectively, $L$ is the fibre length, $\Delta\lambda$ is the bandwidth of the quantum channel, $\Delta t$ is the time-filtering coefficient of the detector, and $\beta_{US}$ is the Raman scattering coefficient. Backward Raman scattering of the downstream signal is given by

$$P_b = P_{DS}\beta_{DS}\Delta\lambda\Delta t \frac{\left(1-e^{-(\alpha_{DS}+\alpha_q)L}\right)}{\alpha_{DS}+\alpha_q} ,$$

where $P_{DS}$ is the launch power, $\alpha_{DS}$ is the fibre attenuation coefficient for 1490 nm, and $\beta_{DS}$ is the Raman scattering coefficient. On both the receiver and transmitter side a total of three filters are necessary to separate quantum and GPON signals and to reduce Raman noise. All three filters are standard telecom filters. A first GPON filter separates QKD and clock signal from the data signals. This is followed by a coarse wavelength division multiplexing (CWDM) filter which separates the clock and quantum signals. As a last step, narrow filtering of the quantum channel is essential to reduce Raman noise. To this end, we integrate matching 25 GHz dense wavelength division multiplexing filters (DWDM) based on fibre Bragg gratings in transmitter and receiver. Please note that the 1550nm wavelength channel used for the QKD signal is reserved in recommendation ITU-T G.984.5 as an enhancement band for additional services such as video-on-demand. Quantum key transmission is another form of additional service and therefore does not conflict with the recommendation.

## *Simulation and secure key rate*

The numerical simulation is based on a noise estimation using experimentally measured parameters. Raman noise counts are calculated in the separate sections of the network using the formulas above and $\beta_{US} = 8 \times 10^{-10}$ nm$^{-1}$, $\beta_{DS} = 6.8 \times 10^{-9}$ nm$^{-1}$, $\alpha_q = 0.046$ km$^{-1}$, $\alpha_{US} = 0.076$ km$^{-1}$, $\alpha_{DS} = 0.051$ km$^{-1}$, $\Delta\lambda = 0.14$ nm, and $\Delta t = 0.127$. Raman noise from the clock signal is also taken into account with $\beta_{clk} = 2.4 \times 10^{-9}$ nm$^{-1}$ and $\alpha_{clk} = 0.051$ km$^{-1}$. The insertion loss of the power splitters is 9.2 dB, 12.7 dB, 16.3 dB, 19.6 dB, and 22.8 dB for 1 × 8, 1 × 16, 1 × 32, 1 × 64, and 1 × 128, respectively. The insertion loss of Bob including GPON, CWDM and FBG filter is 5.5 dB (narrow FBG filter 2.5 dB). Optical errors from modulation and interferometers contribute approximately 0.8% to the QBER. The key rate is secure against collective attacks[31] and is negligibly different from the most general attack available to the eavesdropper[31]. The epsilon parameter is set to $10^{-10}$ and the error correction efficiency is 1.1 times the Shannon limit. We only use bits transmitted in the majority *Z* basis for the final secure key. To compensate the reduction of the secure key rate from finite size effects we increase the key session time from 20 min for 1 × 8 splitter to 30 min, 60 min, 120 min, and 240 min for 1 × 16, 1 × 32, 1 × 64, and 1 × 128 splitter, respectively.